\documentclass{PoS}

\newcommand{\ApJ}{ApJ}

\newcommand{\etal}{et alii}

\newcommand{\AMS}{\textsf{AMS}}

\def\Journal#1#2#3#4{{#4}, {#1}, {#2}, #3} 
\newcommand{\citep}{\cite}

\newcommand{\p}{\textrm{\ensuremath{p}}}

\newcommand{\He}{\textrm{He}}
\newcommand{\Li}{\textrm{Li}}
\newcommand{\Be}{\textrm{Be}}
\newcommand{\B}{\textrm{B}}
\newcommand{\C}{\textrm{C}}
\newcommand{\N}{\textrm{N}}
\newcommand{\Oxy}{\textrm{O}}
\newcommand{\Si}{\textrm{Si}}
\newcommand{\Fe}{\textrm{Fe}}

\newcommand{\BC}{\textrm{B}/\textrm{C}}

\newcommand{\Htwo}{\ensuremath{^{2}}\textrm{H}}
\newcommand{\Het}{\ensuremath{^{3}}\textrm{He}}

\newcommand{\pbarp}{\textrm{\ensuremath{\bar{p}/p}}}
\newcommand{\pbar}{\textrm{\ensuremath{\bar{p}}}}
\newcommand{\nbar}{\textrm{\ensuremath{\bar{n}}}}
\newcommand{\dbar}{\ensuremath{\rm \overline{d}}}

\newcommand{\hebar}{\ensuremath{\rm \overline{He}}}

\newcommand{\hetbar}{\ensuremath{\rm \overline{^{3}He}}}
\newcommand{\htbar}{\ensuremath{\rm \overline{^{3}H}}}

\title{Production of cosmic-ray antinuclei in the Galaxy and background for dark matter searches}
\ShortTitle{Production of cosmic-ray antinuclei in the Galaxy and background for dark matter searches}
\author{{Nicola Tomassetti}\thanks{E-mail: {nicola.tomassetti@cern.ch}}\\
  Department of Physics and Earths Science, Universit{\`a} di Perugia, and INFN-Perugia, I-06100 Perugia, Italy\\
}
\author{{Alberto Oliva}\thanks{E-mail: {alberto.oliva@cern.ch}}\\
Centro de Investigaciones Energ{\'e}ticas, Medioambientales y Tecnol{\'o}gicas CIEMAT, E-28040 Madrid, Spain\\
}

\abstract{
  Antimatter nuclei in cosmic rays (CR) represent a promising discovery channel for the indirect search of dark matter.
  We present astrophysical background calculations of CR antideuteron (\dbar) and antihelium (\hebar). 
  These particles are produced by high-energy collisions of CR protons and nuclei with the gas nuclei of the interstellar medium.
  In our calculations, we also consider production and shock acceleration of antinuclei in the shells of supernova remnants (SNRs).
  The total flux of \dbar{} and \hebar{} particles is constrained using new \AMS{} measurements on the boron/carbon (\BC) and antiproton/proton (\pbarp) ratios.  
  The two ratios leads to different antiparticle fluxes in the high-energy regime $E\gtrsim$\,10\,GeV/n where,
  in particular,  \pbarp-driven calculations leads to a significantly larger antiparticle flux in comparison to predictions from conventional \BC-driven constraints.
  On the other hand, both approaches provide consistent results in the sub-GeV/n energy window,
  which is where dark matter induced signal may exceed the astrophysical background.
  In this region, the \emph{total} antinuclei flux, from interaction in the insterstellar gas and inside SNRs, is tightly bounded by the data.
  Shock-acceleration of antiparticles in SNRs has a minor influence in the astrophysical background for dark matter searches.
}

\FullConference{The European Physical Society Conference on High Energy Physics\\
   5-12 July 2017\\
    Venice, Italy}

\begin{document}

%%%%%%%%%%%%%%%%%%%%%%%%%%%%%%%%%
\section{Introduction}
%%%%%%%%%%%%%%%%%%%%%%%%%%%%%%%%%
Antiproton (\pbar), antideuteron (\dbar), and antihelium (\hebar) are unique messengers for the search DM annihilation signals in the Galaxy.
The antiproton/proton (\pbarp) ratio CRs has been recently measured with high precision 
by the Alpha Magnetic Spectrometer (\AMS) from 0.5 to 450\,GeV of kinetic energy \citep{Aguilar2016PbarP}. 
The detection of antinuclei \dbar{} and \hebar{} is the next milestone in CR physics.
DM annihilation processes into \dbar{} and \hebar{} particles may generate an observable excess in their spectrum \citep{Aramaki2016,Cirelli2014,Fornengo2017}. 
In particular, at energy below a few GeV, the astrophysical background is kinematically suppressed 
while DM signals peak in this energy window according to several models of DM annihilation.
In this paper, we report calculations of secondary antinuclei fluxes and their uncertainties.

%%%%%%%%%%%%%%%%%%%%%%%%%%%%%%%%%%%%%%%%%%%%%%%%%%%%%%%
\section{Calculations}
%%%%%%%%%%%%%%%%%%%%%%%%%%%%%%%%%%%%%%%%%%%%%%%%%%%%%%%
Secondary antinuclei are generated by collisions of CRs with the interstellar matter (ISM) \citep{Grenier2015}. 
Using cross-sections presented in \cite{Tomassetti2012Iso,TomassettiFeng2017,Tomassetti2015XS},
we calculated the production of several isotopes such as  \Htwo, \Het, $^{6,7}$\Li, $^{7,9,10}$\Be, and $^{10,11}$\B{}
from fragmentation of \C-\N-\Oxy, \Si, and \Fe. 
We implemented dedicated algorithms to compute the \pbar{} and \nbar{} production from
\p-\p, \p-\He, \He-\p, and \He-\He{} collisions \cite{DiMauro2014}.
Then, we used an improved nuclear coalscence model to evaluate
the production of \dbar=\{\nbar,\pbar\} and \hebar{} \citep{TomassettiOliva2017}.
Antihelium production includes \htbar=\{\nbar,\nbar,\pbar\} and \hetbar=\{\nbar,\pbar,\pbar\}.
The SNR injection spectrum of primary CRs is calculated from the diffusive-shock-acceleration equation.
Following earlier calculations \citep{TomassettiDonato2012,TomassettiFeng2017,Herms2017},
we also account for the production and destruction of secondary nuclei and antinuclei in SNRs. 
The secondary production rate inside SNRs is regulated by the product $\tau_{\rm snr}n_{-}$ between SNR age and upstream gas density.
To describe CR propagation in the Galaxy, we use a two-halo model of CR diffusion and interactions, which
describes well the spectral hardening of primary CRs \citep{Tomassetti2015TwoHalo,GuoJin2016}.
Near the disk  ($|z|<l\cong$\,0.5\,kpc), the rigidity ($R=p/Z$) dependent diffusion coefficient 
is $K\propto \beta K_{0}(R/{\rm GV})^{\delta_{0}}$, with $\delta_{0}=1/3$.
Away from the disk ($l<|z|<L\cong$\,5\,kpc), the diffusion exponent is $\delta_{0}+\Delta$, where 
the value $\Delta=0.55$ is determined using primary CR spectra \citep{Feng2016}. 
The secondary production rate in the ISM depends on the ratio $L/K_{0}$, which is a free parameter.
The near-Earth CR fluxes are affected by solar modulation, which we have modeled within the \textit{force-field} approximation
using $\phi_{\rm Fisk}=0.7$\,GV as modulation potential for \AMS{} \citep{Ghelfi2016}. 
To constrain the key parameters, we use the \AMS{} data on \BC{} ratio and \pbarp{} ratio.
\BC-driven and \pbarp-driven fits lead to different parameter values, inconsistent each other,
and thus different predictions for the spectra of \dbar{} and \hebar{} \citep{TomassettiOliva2017}.
In the following, \BC-driven predictions are presented.

%%%%%%%%%%%%%%%%%%%%%%%%%%%%%%%%%%%%%%%%%%%%%%%%%%%
\section{Results and discussions}               %%%
\label{Sec::ResultsAbar}                        %%%
%%%%%%%%%%%%%%%%%%%%%%%%%%%%%%%%%%%%%%%%%%%%%%%%%%%

%%%%%%%%%%%%%%%%%%%%%%%%%%%%%%%%%%%%%%%%%%%%%%%%%%%%%%%%%%%%%%%%%%%%%
\begin{figure}[!t] 
\centering
\includegraphics[width=0.95\textwidth]{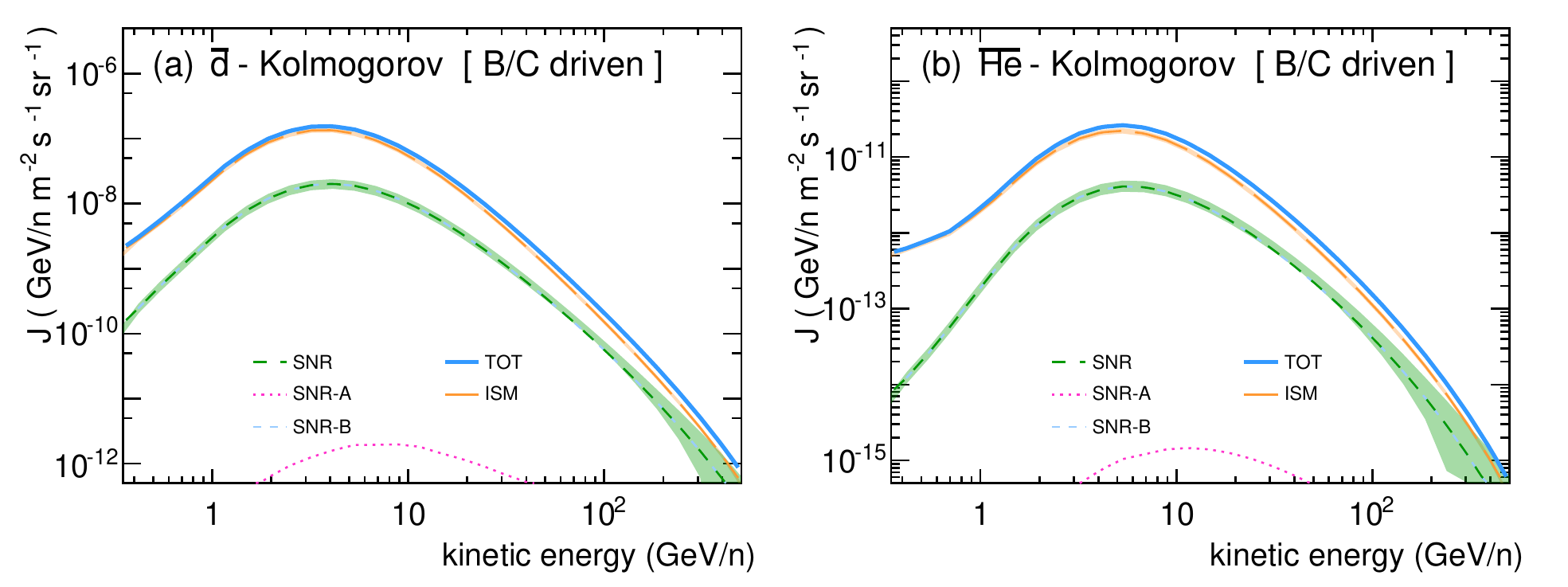}
\caption{\footnotesize%
  Model predictions of for the total flux of antideuterons (a) and antihelium (b), including secondary production inside SNRs 
  (green short-dashed lines), standard production in the ISM (orange long-dashed lines), and total flux (light-blue solid lines).
  Calculations are fromx the \BC-driven parameter setting.
}\label{Fig::ccAbarBCDriven}
\end{figure}
%%%%%%%%%%%%%%%%%%%%%%%%%%%%%%%%%%%%%%%%%%%%%%%%%%%%%%%%%%%%%%%%%%%%%

Our \BC-driven predictions for the \dbar{} and \hebar{} spectra are shown in Fig.\,\ref{Fig::ccAbarBCDriven}.
Our calculations are in agreement with those reported in early works \citep{Donato2008}.
Along with the standard ISM-induced component (orange line),  the SNR-accelerated fluxes (green line) are shown.
As shown in \citep{TomassettiOliva2017}, \pbarp-driven calculations lead to stronger SNR-production that dominates the flux at $E\gtrsim$\,100\,GeV/n.
However, as discussed, \pbarp-driven calculations overproduce the \BC{} ratio.
Interestingly, both \BC-driven and \pbarp-driven calculations lead to consistent results at sub-GeV/n energy.
In spite of considerable errors for the single SNR and ISM component, the \emph{total} ISM+SNR flux prediction
is found to be highly stable for a large region of parameter space.
In this respect, the total astrophysical background for DM searches is well assessed at mid-low energies.
We also recall that the antinuclei flux calculations are affected by large nuclear uncertainties that we have not addressed in this work.
These uncertainties have a similar influence on ISM and SNR components, being the two contributions tightly correlated each other.

%%%%%%%%%%%%%%%%%%%%%%%%%
\section{Conclusions} %%%
%%%%%%%%%%%%%%%%%%%%%%%%%
%
Using new data from \AMS{} on the \BC{} and \pbarp{} ratios, we have presented calculations for the fluxes of \dbar{} and \hebar{} in CRs.
We have compared \BC-driven and \pbar-driven calculations under different transport models \citep{TomassettiOliva2017}.
In the sub-GeV energy window, where DM-induced signatures may exceed the background,
we found that the \emph{total} flux of secondary antinuclei (from CR+gas collisions in the ISM and in SNRs)
is highly stable for a large region of parameter configuration.
Furthermore, at these energies, the SNR-accelerated flux is sub-dominant.
In summary, the astrophysical background for DM searches in the low-energy region appear to be soundly assessed.
The search of \dbar{} and \hebar{} in CRs is ongoing by the \AMS{} experiment, and soon, by the GAPS detection project \citep{GAPS}.
\\
\\
{\footnotesize%
AO acknowledges CIEMAT, CDTI and SEIDI MINECO under grants ESP2015-71662-C2-(1-P) and MDM-2015-0509.
NT acknowledges support from the MAtISSE project.
This project has received funding from the European Union's Horizon 2020 research and innovation programme under the Marie Sklodowska-Curie grant agreement No 707543.
}

\end{document}